\documentclass[twocolumn]{aastex701}
\usepackage{natbib}
\usepackage{amsmath}

\bibpunct{(}{)}{;}{a}{}{,}

\begin{document}

\title{Searching for Isolated Black Hole Candidates within 15~pc of the Solar System in Gaia~DR3}

\author[0009-0009-0287-2535]{Abdurakhmon Nosirov}
\affiliation{Center for Astronomy and Astrophysics, Department of Physics, Fudan University, Shanghai 200438, China}
\affiliation{Institut f\"ur Astronomie und Astrophysik, Eberhard-Karls Universit\"at T\"ubingen, D-72076 T\"ubingen, Germany}
\email{abdurakhmonn24@m.fudan.edu.cn}  

\author[0000-0002-3180-9502]{Cosimo Bambi}
\altaffiliation{bambi@fudan.edu.cn}
\affiliation{Center for Astronomy and Astrophysics, Department of Physics, Fudan University, Shanghai 200438, China}
\affiliation{School of Humanities and Natural Sciences, New Uzbekistan University, Tashkent 100001, Uzbekistan}
\email{bambi@fudan.edu.cn} 

\author[0000-0003-3319-166X]{Leda Gao}
\affiliation{Center for Astronomy and Astrophysics, Department of Physics, Fudan University, Shanghai 200438, China}
\email{ledagao@fudan.edu.cn} 

\author[0000-0001-6459-8599]{Jos de Bruijne}
\affiliation{Directorate of Science, European Space Research and Technology Centre, NL-2201 AZ Noordwijk, The Netherlands}
\email{Jos.De.Bruijne@esa.int}

\author[0000-0002-9639-4352]{Jiachen Jiang}
\affiliation{Department of Physics, University of Warwick, Coventry CV4 7AL, UK}
\email{jcjiang12@outlook.com}

\author[0000-0003-4187-9560]{Andrea Santangelo}
\affiliation{Institut f\"ur Astronomie und Astrophysik, Eberhard-Karls Universit\"at T\"ubingen, D-72076 T\"ubingen, Germany}
\affiliation{Center for Astronomy and Astrophysics, Department of Physics, Fudan University, Shanghai 200438, China}
\email{andrea.santangelo@uni-tuebingen.de}

\author[0000-0001-9969-2091]{Fu-Guo Xie}
\affiliation{Shanghai Astronomical Observatory, Chinese Academy of Sciences, Shanghai 200030, China}
\email{fgxie@shao.ac.cn}

\begin{abstract}
Theoretical models predict that the Galaxy hosts $10^8$-$10^9$ black holes formed from the complete gravitational collapse of heavy stars and that most of these black holes are isolated, without any companion. Within 15~pc of the Solar System ($\sim 50$~ly), there may be a few black holes. If located inside one of the Local Interstellar Clouds -- which occupy 5-20\% of this local volume -- an isolated black hole could produce detectable electromagnetic emission via accretion from the interstellar medium, given the capabilities of current or near-future observatories. However, precise predictions remain challenging due to large uncertainties in the expected accretion spectra. Outside these clouds, the accretion rate would be too low; according to our models, the resulting electromagnetic flux is well below the detection thresholds of current and near-future observational facilities. While astrometric detection via gravitational perturbation of nearby stars is conceivable, the local stellar density is too low for this method to be realistically successful. We have searched the Gaia~DR3 catalog for candidate isolated black holes accreting from the interstellar medium and identified five sources. All candidates lie close to the Galactic plane, making them likely spurious astrometric solutions, for instance caused by unmodelled background sources (crowding) and/or unmodelled binarity. Our search for infrared and radio emission from these sources further suggests that they are unlikely to be black holes accreting from the interstellar medium.
\end{abstract}

\section{Introduction}

Black holes represent one of the most intriguing predictions of Einstein's theory of General Relativity and serve as ideal laboratories for testing gravity in the strong-field regime~\citep[e.g.,][]{2017RvMP...89b5001B,2017bhlt.book.....B,2016CQGra..33e4001Y,2024rpgt.book.....B}. Prior to 2015, their observational evidence was indirect, relying on inferences from stellar-mass black holes in X-ray binaries and supermassive black holes in galactic nuclei. The identification of these objects as the Kerr black holes of General Relativity~\citep{1963PhRvL..11..237K,1971PhRvL..26..331C,1975PhRvL..34..905R} was largely due to the absence of alternative explanations within conventional physics: stellar-mass black holes exceeded the maximum theoretical mass of neutron stars~\citep{1974PhRvL..32..324R,2012ARNPS..62..485L}, while supermassive black holes were too massive, compact, and old to be clusters of neutron stars~\citep{1998ApJ...494L.181M}. Today, we can directly probe the nature of these objects through gravitational waves~\citep[e.g.,][]{2016PhRvL.116v1101A,2019PhRvD.100j4036A}, X-ray observations~\citep[e.g.,][]{2018PhRvL.120e1101C,2019ApJ...875...56T,2019ApJ...874..135T,2021ApJ...913...79T}, and black hole imaging~\citep[e.g.,][]{2020PhRvL.125n1104P,2022ApJ...930L..17E,2023CQGra..40p5007V}, confirming the predictions of General Relativity within the precision of our measurements.

While the next generation of observational facilities will improve existing constraints, performing very precise and accurate tests of General Relativity with astrophysical observations may remain difficult due to the complex astrophysical environments around black holes. This motivates consideration of alternative approaches, such as sending a probe to orbit a nearby black hole~\citep{2025iSci...28k3142B,2025arXiv250911222B}. Although extremely challenging, such a mission -- employing a small spacecraft or fleet of small spacecraft -- is not inconceivable in the coming decades. A critical requirement is that the target black hole lie within 15~pc (approximately 50~ly) of the Solar System; at greater distances, a complete mission profile -- including travel, operations, and data return -- would exceed a century even for a spacecraft traveling near the speed of light.

Motivated by this idea, in this work we investigate the possibility of discovering an isolated black hole within 15~pc of the Solar System. As first noted by \citet{1971SvA....15..377S}, an isolated black hole can accrete from the interstellar medium, potentially producing detectable electromagnetic radiation. Subsequent studies have explored this topic \citep{1975A&A....44...59M,1985MNRAS.217...77M,1998ApJ...495L..85F,2005MNRAS.360L..30M,2013MNRAS.430.1538F,2018MNRAS.477..791T,2021MNRAS.505.4036S,2025ApJ...988L..12M}, though none has attempted to identify candidates from existing astronomical observations.

Within 15~pc of the Solar System, 80-95\% of the volume is filled with a hot, low-density interstellar medium~\citep{2008ApJ...673..283R}. As shown in the following section, the particle density in this region is too low for accretion onto an isolated black hole to produce a luminosity above the detection threshold of current or near-future facilities, regardless of the accretion model. The situation changes if a black hole resides within one of the Local Interstellar Clouds, where the interstellar medium is warm, partially ionized, and denser. These clouds occupy only 5-20\% of the local volume, reducing the probability of finding a black hole inside them. Nevertheless, as discussed in subsequent sections, there are reasonable prospects for detecting such objects -- should they exist -- in the near future, despite large uncertainties in their expected accretion spectra.

We therefore search the Gaia~DR3 catalog for candidate isolated black holes accreting from the interstellar medium and identify five sources. All lie close to the Galactic plane, making them likely spurious astrometric solutions. Our failure to detect infrared or radio emission from these sources reinforces the conclusion that they are unlikely to be accreting black holes. However, they cannot be ruled out as genuine black holes without further observations. We also consider the possibility of indirectly detecting non-accreting isolated black holes through their gravitational influence on the motions of nearby stars. However, the low stellar density in the Solar neighborhood makes close encounters between a black hole and a star extremely unlikely on human timescales.

This manuscript is organized as follows. In Section~\ref{s-population}, we review estimates of the black hole population within 15~pc of the Solar System. In Section~\ref{s-spectra}, we discuss accretion from the interstellar medium onto an isolated black hole and compare predicted spectra with detection thresholds of current catalogs. Our search for candidates in Gaia~DR3 is presented in Section~\ref{s-gaia}. A summary and our conclusions are given in Section~\ref{s-conclusions}.


\section{Population of Nearby Black Holes}\label{s-population}

Estimates of the population of black holes in the Solar neighborhood are subject to significant uncertainties, encompassing their number, local environments, and expected spectra. In this section, we review and elaborate on previous estimates of the expected abundance of black holes within 15~pc of the Solar System. We conclude that it is plausible for a few such objects to reside within this volume.

Theoretical models suggest the Galaxy contains between $\sim 10^8$~\citep{2020A&A...638A..94O} and $\sim 10^9$~\citep{1996ApJ...457..834T} stellar-mass black holes ($N_{\rm BH}$). Most are expected to be isolated, having lost any companion star during their evolution. Although most massive stars form in binary systems, the binary survival probability is low: systems with small separations may merge during the red supergiant phase of the massive star, while wider systems are typically disrupted by the supernova explosion that produces the black hole. \citet{2020A&A...638A..94O} estimate that the Galactic disk hosts $\sim 1.0 \cdot 10^8$ isolated black holes and $\sim 8 \cdot 10^6$ black holes in binary systems; of the latter, about 80\% have another black hole as a companion, and fewer than 2\% are paired with a normal star.

To estimate the proximity of the nearest black hole, we consider two approaches. \citet{2013MNRAS.430.1538F} model the Galaxy as a disk (radius 10~kpc, thickness 0.5~kpc) plus a bulge (radius 2~kpc), yielding a total volume $V_{\rm Gal} \sim 200$~kpc$^3$. The average volume per black hole is therefore $V_{\rm BH} = V_{\rm Gal}/N_{\rm BH}$, ranging from 200~pc$^3$ (for $N_{\rm BH} = 10^9$) to 2,000~pc$^3$ (for $N_{\rm BH} = 10^8$). If we model this volume as a sphere, its radius ranges from 3.6~pc to 7.8~pc, respectively. Placing the Solar System at the center of such a sphere implies the nearest black hole lies within 3.6-7.8~pc, with an expected distance in the range 2.9-6.2~pc (9.5-20~ly).

\citet{2025ApJ...988L..12M} adopt a different method, based on the total stellar mass of the Galaxy, $M_{\rm star} = (6.1 \pm 1.1) \cdot 10^{10}~M_\odot$. The ratio $M_{\rm star}/N_{\rm BH}$ then ranges from $\sim 60~M_\odot$ (for $N_{\rm BH}=10^9$) to $\sim 600~M_\odot$ (for $N_{\rm BH}=10^8$). With a local stellar mass density $\rho_{\rm star} \sim 0.04~M_\odot$~pc$^{-3}$~\citep{2025ApJ...990...88L}, the average volume per black hole near the Sun is $V_{\rm BH} = (M_{\rm star}/N_{\rm BH})/\rho_{\rm star}$, or 1,500-15,000~pc$^3$. Modeling this as a sphere gives a radius of 7-15~pc. The expected distance to the nearest black hole would thus be 5.6~pc (18~ly) if within 7~pc, or 12~pc (40~ly) if within 15~pc.

Combining these estimates, the nearest black hole may lie at $\sim 10$-20~ly (from \citealt{2013MNRAS.430.1538F}) or $\sim 20$-40~ly (from \citealt{2025ApJ...988L..12M}). However, the latter approach may overestimate the distance, as the local stellar mass density $\rho_{\rm star}$ does not account for remnant mass (e.g., neutron stars and black holes) or ejected material. Since the Solar neighborhood lacks very young stars, the original massive stars have already evolved, leaving a lower present-day $\rho_{\rm star}$ than in star-forming regions, potentially biasing the inferred black hole distance upward.

The location of the Solar System within a spiral arm may further influence these estimates. Indeed, it is remarkable that all known black hole low-mass X-ray binaries are located in spiral arms~\citep[e.g.,][]{2025MNRAS.541..553A}. Unlike black hole high-mass X-ray binaries, which must be young systems because their companion stars are massive and have correspondingly short lifetimes, black hole low-mass X-ray binaries can be very old systems. This fact is unlikely to be coincidental, even though spiral arms may be transient structures. If such a spatial preference also holds for isolated black holes, we should expect a higher black hole density in spiral arms. Assuming for simplicity that black holes are confined to spiral arms (which occupy $\sim 20\%$ of the Galactic disk volume), the effective volume $V_{\rm Gal}$ in the model of \citet{2013MNRAS.430.1538F} reduces to $\sim 40$~kpc$^3$. The average volume per black hole then becomes 40–400~pc$^3$, and the expected distance to the nearest black hole drops to 1.7–3.6~pc (5.5–12~ly).

Applying a similar correction to the approach of \citet{2025ApJ...988L..12M}, we assume a stellar mass in the spiral arms of $M_{\rm star} \sim 1.5 \cdot 10^{10}~M_\odot$ and negligible black holes elsewhere. Then $V_{\rm BH}$ ranges from 400~pc$^3$ to 4,000~pc$^3$, implying the nearest black hole lies within 4.6-9.8~pc, with an expected distance of 3.7-7.8~pc (12-25~ly). Thus, if black holes are concentrated in spiral arms, the nearest one could be as close as $\sim 6$-12~ly (per \citealt{2013MNRAS.430.1538F}) or $\sim 12$-25~ly (per \citealt{2025ApJ...988L..12M}). While this scenario likely overestimates the local black hole density, it suggests that the true density may be higher than inferred from a smooth Galactic model.


\section{Spectra of Nearby Black Holes Accreting from the Interstellar Medium}\label{s-spectra}

Any isolated black hole in the Galaxy is surrounded by the interstellar medium, enabling accretion that may produce detectable electromagnetic radiation~\citep{1971SvA....15..377S,1975A&A....44...59M,1985MNRAS.217...77M,1998ApJ...495L..85F,2005MNRAS.360L..30M,2013MNRAS.430.1538F,2018MNRAS.477..791T,2021MNRAS.505.4036S,2025ApJ...988L..12M}.

\subsection{Mass Accretion Rate}

The standard framework for describing accretion from the interstellar medium onto an isolated compact object is the Bondi-Hoyle-Lyttleton (BHL) model~\citep{1939PCPS...35..405H,1944MNRAS.104..273B}. In this model, the accretion rate for a black hole of mass $M_{\rm BH}$ moving at velocity $v_{\rm BH}$ relative to the interstellar medium is 
\begin{eqnarray}
\dot{M}_{\rm BHL} = 
\frac{4 \pi G_{\rm N}^2 M_{\rm BH}^2 \, \rho}{\left( v_{\rm BH}^2 + c_{\rm s}^2 \right)^{3/2}} \, ,
\end{eqnarray}
where $G_{\rm N}$ is Newton’s gravitational constant, $c_{\rm s}$ is the sound speed in the interstellar medium, and $\rho$ is the interstellar medium density. The density can be expressed as $\rho = \mu n m_p$, where $\mu$ is the mean atomic mass, $n$ is the particle number density, and $m_p$ is the proton mass. To account for outflows and convection, a dimensionless suppression factor $\lambda < 1$ is often introduced, giving the effective accretion rate as
\begin{eqnarray}\label{eq-lambda}
\dot{M}_{\rm BH} = \lambda \, \dot{M}_{\rm BHL} \, .
\end{eqnarray}

The numerical simulations reported by \citet{2023ApJ...950...31K} and \cite{2025ApJ...978..148G} suggest $\lambda \approx 0.2$-0.5. However, the value of $\lambda$ is highly sensitive to $v_{\rm BH}$. Although the velocity distribution of nearby isolated black holes is unknown, two extreme cases can be considered~\citep{2013MNRAS.430.1538F}. The velocity dispersion of nearby massive stars ($M > 2~M_\odot$) is relatively well known, with $v_\star \sim 20$~km/s~\citep{1998MNRAS.298..387D}; this can be taken as a lower limit for typical values of $v_{\rm BH}$. If black holes receive a kick during the supernova explosion of the progenitor star, higher velocities are expected. Assuming that black holes and neutron stars receive comparable kicks, we can rescale the velocity distribution measured for neutron stars to account for the mass difference, yielding $v_{\rm BH} \sim 60$~km/s \citep{2013MNRAS.430.1538F}.

\begin{figure}[t]
\centering
\includegraphics[width=0.42\textwidth]{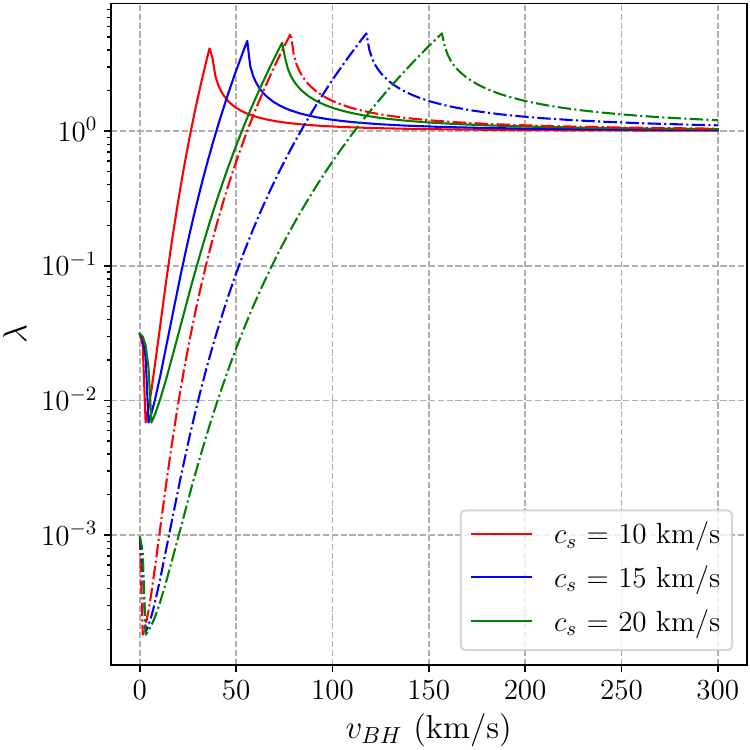}
\caption{Predictions for the phenomenological parameter $\lambda$ in Eq.~(\ref{eq-lambda}) from \citet{2013ApJ...767..163P}. The red, blue, and green curves correspond to $c_{\rm s} = 10$, 15, and 20~km/s, respectively. Solid curves represent $c_{\rm ion}/c_{\rm s} = 2$ and dash-dotted curves represent $c_{\rm ion}/c_{\rm s} = 4$, where $c_{\rm ion}$ is sound speed in the ionized medium surrounding the black hole.}
\label{f-lambda}
\end{figure}

\citet{2013ApJ...767..163P} compared numerical simulations with BHL model predictions and found that the mass accretion rate in their simulations is significantly lower than the BHL prediction for low $v_{\rm BH}$ ($\lambda \ll 1$), higher for ``intermediate'' velocities ($\lambda > 1$), and in good agreement for high $v_{\rm BH}$ ($\lambda \approx 1$); Fig.~\ref{f-lambda} shows the value of $\lambda$ from to the analysis of \citet{2013ApJ...767..163P}. These results are consistent with, for example, the case of Sgr~A$^*$, which is not moving through the interstellar medium: the BHL accretion rate for Sgr~A$^*$ is estimated to be $\dot{M}_{\rm BHL} \approx 3 \times 10^{-5}$~$M_\odot$~yr$^{-1}$~\citep{1999ApJ...517L.101Q}, while its actual accretion rate is $\dot{M}_{\rm BH} \approx 9 \times 10^{-9}$~$M_\odot$~yr$^{-1}$~\citep{2022ApJ...930L..16E}, implying $\lambda \approx 0.0003$.

While numerous studies report results from specific simulations, those results cannot be directly generalized to estimate mass accretion rates for different systems. In contrast, the Park-Ricotti (PR) model~\citep{2013ApJ...767..163P} provides an effective BHL-based formula that enables straightforward calculation of the actual mass accretion rate for any system. In the PR model, the accretion rate is
\begin{eqnarray}
\dot{M}_{\rm PR} = 
\frac{4 \pi G_{\rm N}^2 M_{\rm BH}^2 \, \rho_{\rm ion}}{\left( v_{\rm ion}^2 + c_{\rm ion}^2 \right)^{3/2}} \, ,
\end{eqnarray}
where $\rho_{\rm ion}$, $v_{\rm ion}$, and $c_{\rm ion}$ are the density, relative velocity, and sound speed in the ionized medium surrounding the black hole; these are related to the corresponding interstellar medium quantities as described in \citet{2013ApJ...767..163P}. The PR model recovers the BHL rate at high $v_{\rm BH}$, predicts higher accretion at intermediate velocities, and predicts significantly lower accretion at low velocities \citep[see Fig.~\ref{f-lambda} and][]{2021MNRAS.505.4036S}. Unlike the BHL model, the PR formulation does not require an additional suppression factor $\lambda$, as such effects are already incorporated. However, the PR model has its own limitations: it is a 2D axisymmetric model, which may miss important 3D effects (e.g., instabilities and anisotropic radiation feedback); it assumes that the radiation near the black hole is emitted isotropically, whereas the formation of optically thick structures near the black hole could change the angular dependence of the emission; the simulations in \citet{2013ApJ...767..163P} assume a surrounding medium density $> 10^2$~cm$^{-3}$, while in our case the interstellar medium density is lower, so the accretion process is likely less affected by radiation feedback.

Within the spherical volume of radius 15~pc around the Solar System ($\sim 14,000$~pc$^3$), 80-95\% of the volume is filled with a hot, low-density interstellar medium. The remaining 5-20\% is occupied by a complex of warm, partially ionized clouds known as the Cluster of Local Interstellar Clouds~\citep{2008ApJ...673..283R}. Other interstellar medium phases are negligible in volume and thus unlikely to host an isolated black hole. In the hot, low-density phase, the particle density is $n \sim 0.005$-0.05~cm$^{-3}$, the mean atomic mass is $\mu \approx 0.5$, and the sound speed is $c_{\rm s} \sim 200$~km~s$^{-1}$ (so $c_{\rm s}^2 \gg v_{\rm BH}^2$, dominating the accretion rate). In the warm, partially ionized clouds, $n \sim 0.3$~cm$^{-3}$, $\mu \approx 0.75$, and $c_{\rm s} \sim 10$~km~s$^{-1}$ (so $c_{\rm s}^2 \ll v_{\rm BH}^2$, making $v_{\rm BH}$ the main determinant of accretion).

\begin{figure}[t]
\centering
\includegraphics[width=0.45\textwidth]{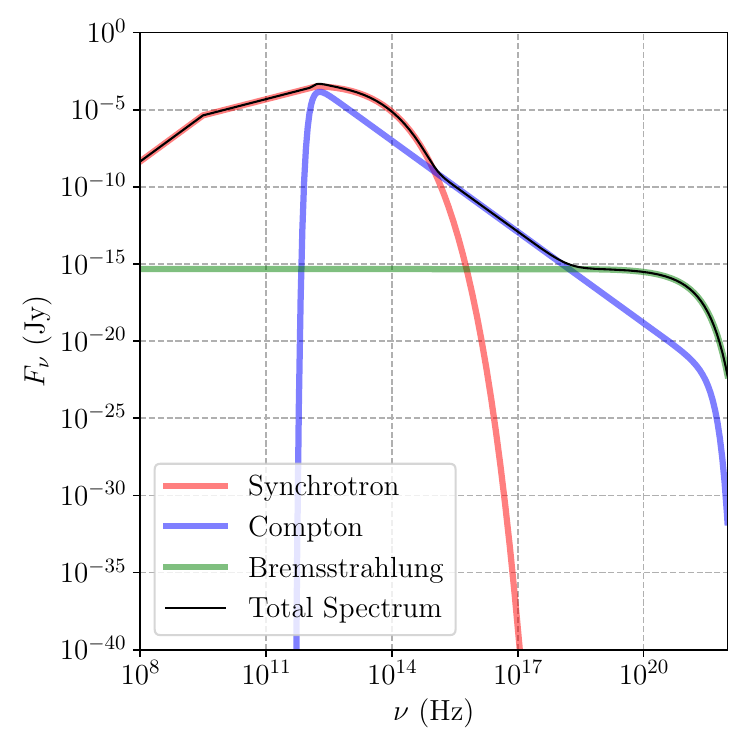}
\caption{Spectrum of an isolated black hole accreting from the interstellar medium as computed with the {\tt LLAGNSED} model. We assume that the mass accretion rate is described by the BHL model with $\lambda = 1$ and we employ the following values for the parameters of the system: black hole mass $M_{\rm BH} = 10$~$M_\odot$, black hole distance $D = 15$~pc, black hole speed $v_{\rm BH} = 50$~km~s$^{-1}$, sound speed in the interstellar medium $c_{\rm s} = 10$~km~s$^{-1}$, particle number density $n = 0.3$~cm$^{-3}$, and mean atomic mass $\mu = 0.75$. The resulting Eddington-scaled accretion luminosity is $\sim 10^{-9}$. The plot shows the three components of the spectrum: synchrotron radiation, inverse Compton scattering, and bremsstrahlung emission.}
\label{f-ADAF}
\end{figure}

\begin{figure*}[htpb]
\centering
\includegraphics[width=0.95\textwidth]{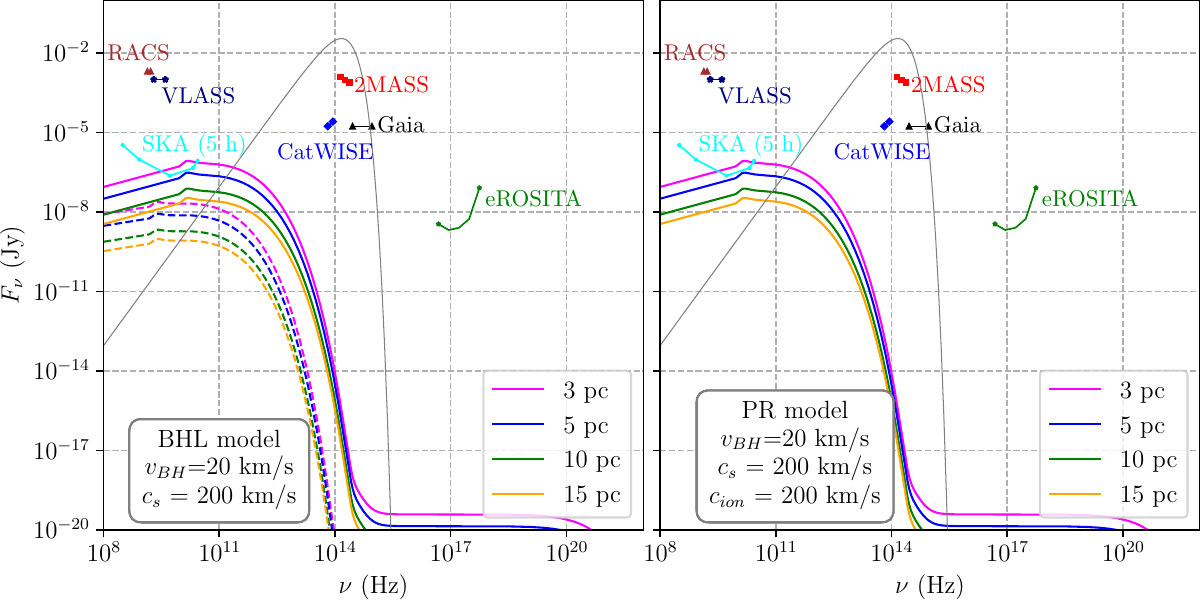}
\caption{Spectra of isolated black holes accreting from a hot and low-density interstellar medium at 3, 5, 10, and 15~pc from the Solar System as computed with {\tt LLAGNSED} (ADAF model). We assume that the particle number density is $n = 0.05$~cm$^{-3}$, the mean atomic mass is $\mu = 0.5$, the sound speed is $c_{\rm s} = 200$~km~s$^{-1}$, the black hole mass is $M_{\rm BH} = 10$~$M_\odot$, and the black hole velocity is $v_{\rm BH} = 20$~km~s$^{-1}$. In the left panel, we assume the BHL accretion model with $\lambda = 1$ (solid) and $\lambda = 0.1$ (dashed). In the right panel, we assume the PR accretion model with $c_{\rm ion} = 200$~km~s$^{-1}$. The gray curve is the spectrum of a brown dwarf with surface temperature $T = 2,500$~K, radius $R = 60,000$~km, at a distance of 10~pc, approximated with a blackbody spectrum. We also report the detection thresholds of Gaia~DR3 (black), eROSITA (green), 2MASS (red), CatWISE (blue), VLASS~QL1 (dark blue), RACS (brown), and a 5~hour observation with SKA (cyan).}
\label{f-hot}
\vspace{0.5cm}
\end{figure*}

\begin{figure*}[htpb]
\centering
\includegraphics[width=0.95\textwidth]{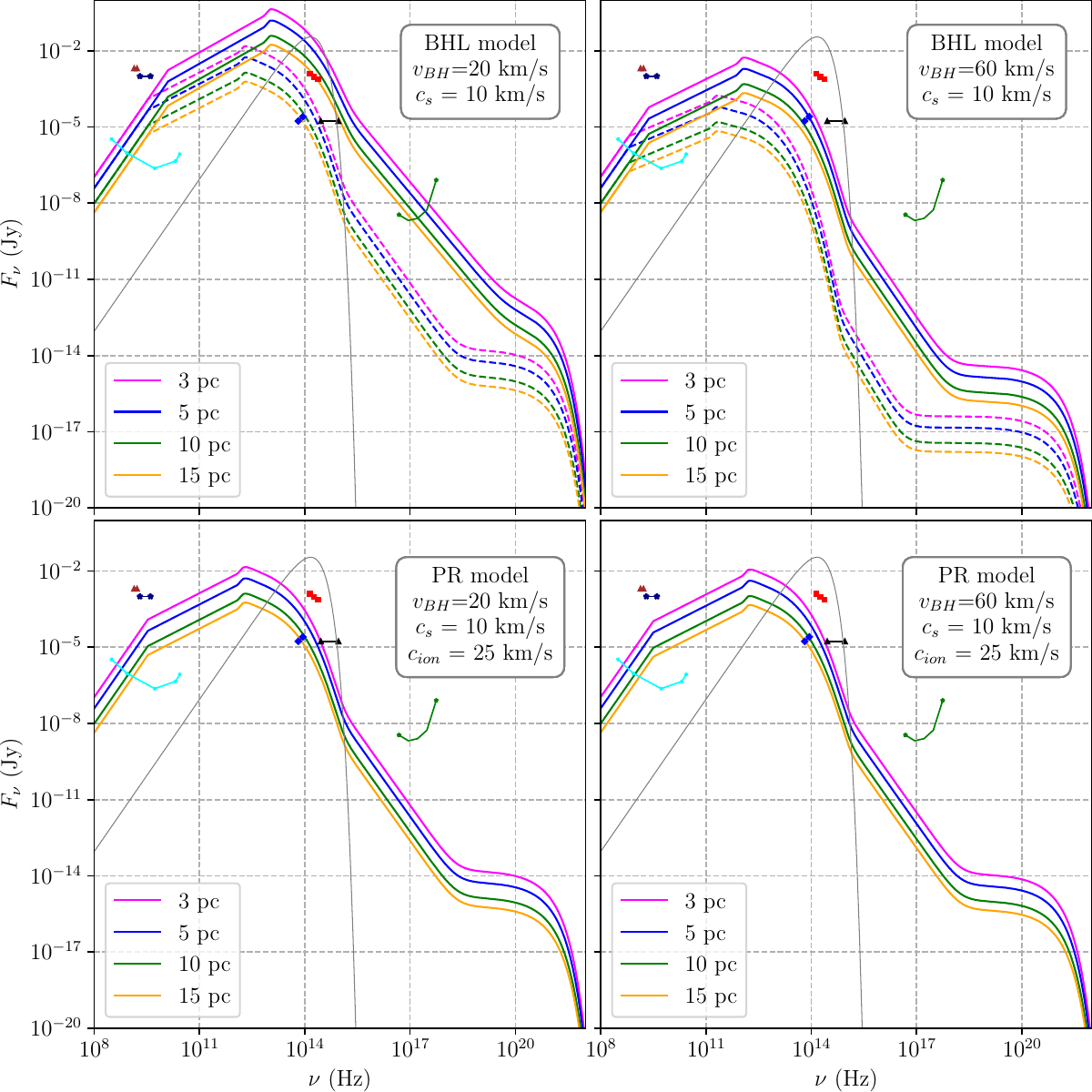}
\caption{Spectra of isolated black holes accreting from a warm and partially ionized interstellar medium at 3, 5, 10, and 15~pc from the Solar System as computed with {\tt LLAGNSED} (ADAF model). We assume that the particle number density is $n = 0.3$~cm$^{-3}$, the mean atomic mass is $\mu = 0.75$, the sound speed is $c_{\rm s} = 10$~km~s$^{-1}$, and the black hole mass is $M_{\rm BH} = 10$~$M_\odot$. In the top panels, we assume the BHL accretion model with $\lambda = 1$ (solid) and $\lambda = 0.1$ (dashed). In the bottom panels, we assume the PR accretion model with $c_{\rm ion} = 25$~km~s$^{-1}$. The black hole velocity is $v_{\rm BH} = 20$~km~s$^{-1}$ in the left panels and $v_{\rm BH} = 60$~km~s$^{-1}$ in the right panels. The gray curve is the spectrum of a brown dwarf with surface temperature $T = 2,500$~K, radius $R = 60,000$~km, at a distance of 10~pc, approximated with a blackbody spectrum. The detection thresholds are the same as in Fig.~\ref{f-hot}.}
\label{f-warm}
\end{figure*}

\begin{figure*}[htpb]
\centering
\includegraphics[width=0.95\textwidth]{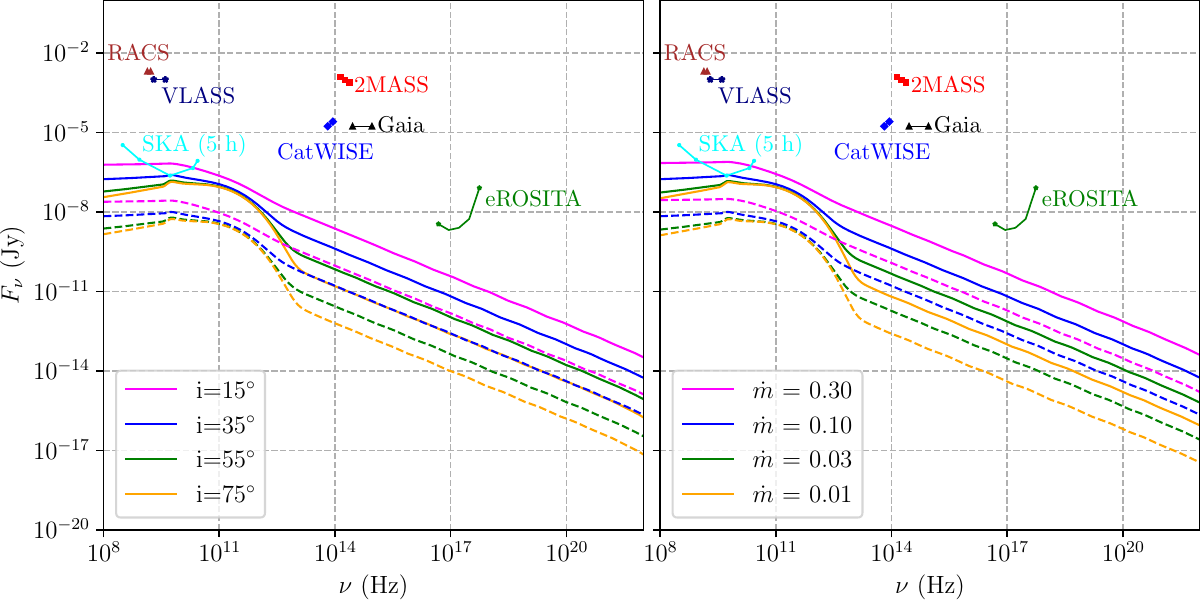}
\caption{Spectra of isolated black holes accreting from a hot and low-density interstellar medium in the ADAF+Jet model. In the left panel, we vary the angle between the jet axis and our line of sight assuming that the mass-loss rate into jet is 10\% of the black hole mass accretion rate. In the right panel, we vary the mass-loss rate into jet assuming that the angle between the jet axis and our line of sight is 35$^\circ$. In both panels, we assume the BHL accretion model with $\lambda = 0.3$ and we employ the following values for the parameters of the system: particle number density $n = 0.05$~cm$^{-3}$, mean atomic mass $\mu = 0.5$, sound speed $c_{\rm s} = 200$~km~s$^{-1}$, black hole mass $M_{\rm BH} = 10$~$M_\odot$, and black hole velocity $v_{\rm BH} = 20$~km~s$^{-1}$. Solid curves are for a black hole distance $D = 3$~pc and dashed curves are for a black hole distance $D = 15$~pc. The detection thresholds are the same as in Fig.~\ref{f-hot}.}
\label{f-hot2}
\vspace{1.0cm}
\centering
\includegraphics[width=0.95\textwidth]{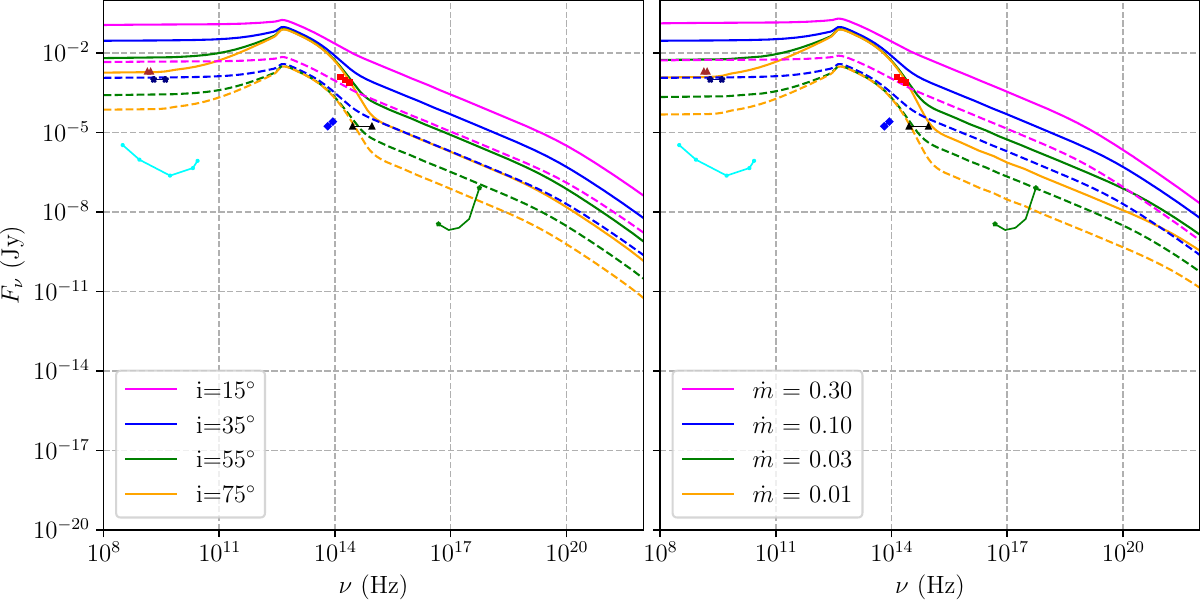}
\caption{Spectra of isolated black holes accreting from a warm and partially ionized interstellar medium in the ADAF+Jet model. In the left panel, we vary the angle between the jet axis and our line of sight assuming that the mass-loss rate into jet is 10\% of the black hole mass accretion rate. In the right panel, we vary the mass-loss rate into jet assuming that the angle between the jet axis and our line of sight is 35$^\circ$. In both panels, we assume the BHL accretion model with $\lambda = 0.3$ and we employ the following values for the parameters of the system: particle number density $n = 0.3$~cm$^{-3}$, mean atomic mass $\mu = 0.75$, sound speed $c_{\rm s} = 10$~km~s$^{-1}$, black hole mass $M_{\rm BH} = 10$~$M_\odot$, and black hole velocity $v_{\rm BH} = 20$~km~s$^{-1}$. Solid curves are for a black hole distance $D = 3$~pc and dashed curves are for a black hole distance $D = 15$~pc. The detection thresholds are the same as in Fig.~\ref{f-hot}.}
\label{f-warm2}
\end{figure*}

\subsection{ADAF Model}\label{ss-adaf}

The accretion rate onto an isolated black hole accreting from the interstellar medium is extremely low. For the BHL model with $\lambda = 1$, the Eddington-scaled luminosity of a black hole inside one of the Local Interstellar Clouds is $L/L_{\rm Edd} \sim 10^{-9}$. Outside these clouds, in the hot, low-density interstellar medium, this value drops below $10^{-17}$. The standard framework for accretion at such low rates is the advection-dominated accretion flow (ADAF) model~\citep{1994ApJ...428L..13N,1995ApJ...444..231N,1995ApJ...452..710N}.

Following \citet{2025ApJ...988L..12M}, we use the public code {\tt LLAGNSED} \citep{2021ApJ...923..260P}, developed for low-luminosity active galactic nuclei but scalable to stellar-mass black holes. We adopt default parameters and set the black hole mass to $10~M_\odot$, noting that the model can become unstable for certain input combinations. Fig.~\ref{f-ADAF} shows a representative spectrum from {\tt LLAGNSED}, which combines synchrotron radiation, bremsstrahlung emission, and inverse Compton scattering. The {\tt LLAGNSED} code originally could not correctly solve Eq.~(A16) in \citet{2021ApJ...923..260P} for the minimum frequency $\nu_{\rm m}$ and the peak frequency $\nu_{\rm p}$, which are critical for calculating synchrotron radiation and inverse Compton scattering. This issue was fixed in the version of {\tt LLAGNSED} released in December 2025 (Pesce, private communication).

Fig.~\ref{f-hot} shows spectra for black holes accreting from the hot, low-density interstellar medium at distances of 3, 5, 10, and 15~pc, computed with {\tt LLAGNSED}. The left panel assumes the BHL model; the right panel assumes the PR model. Even for the highest plausible density in this phase ($n = 0.05$~cm$^{-3}$), all predicted spectra lie well below the detection thresholds of current catalogs. Fig.~\ref{f-hot} includes sensitivity limits for Gaia~DR3~\citep{2023A&A...674A...1G}, eROSITA~DR1~\citep{2024A&A...682A..34M}, 2MASS~\citep{2003tmc..book.....C}, CatWISE~\citep{2021ApJS..253....8M}, VLA Sky Survey Quick Look 1~\citep[VLASS~QL1,][]{2020PASP..132c5001L}, Rapid ASKAP Continuum Survey~\citep[RACS,][]{2024PASA...41....3D,2025PASA...42...38D}, and a 5-hour observation with the Square Kilometre Array~\citep[SKA,][]{2019arXiv191212699B}.

Fig.~\ref{f-warm} shows spectra for black holes accreting from the warm, partially ionized clouds. The top panels show BHL predictions for $\lambda = 1$ (solid) and $\lambda = 0.1$ (dashed); the bottom panels show PR predictions. Left and right panels assume $v_{\rm BH} = 20$ and $60$~km~s$^{-1}$, respectively. The same sensitivity limits as in Fig.~\ref{f-hot} are shown.

In the BHL model, the accretion rate decreases monotonically with $v_{\rm BH}$. Fig.~\ref{f-warm} indicates that a black hole in one of the Local Interstellar Clouds may be above current detection thresholds for $v_{\rm BH} = 20$~km~s$^{-1}$ but fall below them for $v_{\rm BH} = 60$~km~s$^{-1}$. In the PR model, the accretion rate peaks around $v_{\rm BH} \approx 50$~km~s$^{-1}$ for this interstellar medium phase and declines at lower and higher velocities. We caution that the {\tt LLAGNSED} spectra, based on self-similar ADAF solutions with an ad hoc inner cutoff at $6~r_g$ (where $r_g = G_{\rm N} M_{\rm BH}/c^2$), provide a general guide rather than a precise prediction. Altering this inner radius, for example, shifts the synchrotron cutoff to higher energies, which would affect the relative sensitivity of facilities like Gaia and CatWISE.

\begin{table*}[t]
\centering
\renewcommand\arraystretch{1.3}
\caption{Sources within 15~pc of the Solar System in the Gaia~DR3 catalog and without a clear classification. The precision of the position in the sky of these sources is at the level of 1-2~mas in the Gaia~DR3 and is not reported in this table, where we truncate the Galactic coordinates at the second decimal digit. The G magnitude of source 4114677389434590720 is not available in {\tt gaia\_source} but at \href{https://www.cosmos.esa.int/web/gaia/dr3-known-issues\#PhotometryMissingGFluxes}{https://www.cosmos.esa.int/web/gaia/dr3-known-issues\#PhotometryMissingGFluxes}.
}
\label{t-gaia}
\vspace{0.2cm}
\begin{tabular}{|ccccc|}
\hline
Source ID & Galactic Coordinates & Parallax & G~Magnitude & Classification \\
\hline
2021981409373336448 &  $(61.57^\circ , 1.73^\circ)$      & $76.2 \pm 2.8$   & 20.7  & Unknown Source \\
3489874340630661248 &  $(292.04^\circ , 38.33^\circ)$  & $94.2 \pm 1.1$   & 20.6  & Brown Dwarf Candidate \\
4039831777499368960 &  $(352.65^\circ , -4.79^\circ)$   & $81.8 \pm 2.9$   & 20.0  & Unknown Source \\
4114677389434590720 &  $(1.84^\circ , 8.40^\circ)$        & $67.9 \pm 2.7$   & 20.4  & Unknown Source \\ 
4269379774976572160 &  $(38.61^\circ , -1.67^\circ)$     & $84.4 \pm 2.5$   & 20.0  & Unknown Source \\
4318384355378007424 &  $(51.87^\circ , -2.47^\circ)$     & $101 \pm 3$       & 20.6  & Unknown Source \\
\hline
\end{tabular}
\vspace{0.8cm}
\end{table*}

\subsection{ADAF+Jet Model}\label{ss-adaf+jet}

If a system is very faint, with a total luminosity $L < 10^{-6}~L_{\rm Edd}$, some models predict that the spectrum is jet-dominated~\citep{2005ApJ...629..408Y}. Additional observational evidence suggests that black holes accreting at very low rates may indeed be jet-dominated~\citep{2014MNRAS.442L.110X,2016RAA....16...62Y}. This situation certainly applies to isolated black holes accreting from the interstellar medium, given their extremely low mass accretion rates.

We therefore combine the ADAF spectrum from {\tt LLAGNSED} with a jet spectrum using the phenomenological model of \citet{2014MNRAS.442L.110X}. This model adopts the internal shock scenario, widely used in the gamma-ray burst community to interpret afterglow emission. Like all available jet models, it assumes a power-law distribution of density, magnetic field, and other quantities, and follows the spirit of the Blandford–Znajek mechanism~\citep{1977MNRAS.179..433B}. A similar but more complex version can be found in \citet{2014MNRAS.440.2238Z}, which includes a more detailed treatment of the energy distribution near the cutoff and additionally accounts for ``external photon'' radiative cooling via Compton scattering of stellar photons. Since our work focuses on isolated black holes, this additional complexity is not relevant.

The general picture shared by these models is that, at some distance from the black hole (assumed here to be 10 gravitational radii), the Poynting flux jet enters an energy dissipation region, where emission is generated, driven by the tiny mass supply from the accretion disk. The jet power is determined by both the mass-loss rate in the jet and the bulk Lorentz factor. However, these assumptions do not significantly affect the final results, which are consistent with the predictions of other jet models, with the exception of models that include the jet base and require higher baryonic mass loading near the black hole \citep[e.g.,][]{2003A&A...397..645M}.

We assume a mildly relativistic jet with Lorentz factor 1.8, as suggested by the case of quiescent black hole X-ray binaries. The key-model parameters are the mass-loss rate into the jet, $\dot{M}_{\rm jet}$, and the inclination angle $i$ between the jet axis and our line of sight. The jet mass-loss rate is constrained by $\dot{m} \equiv \dot{M}_{\rm jet} / \dot{M}_{\rm BH} < 1$. Details on the model and its parameters can be found in \citet{2005ApJ...620..905Y} and \citet{2016MNRAS.463.2287X}.

Figs.~\ref{f-hot2} and~\ref{f-warm2} show the resulting ADAF+Jet spectra for accretion from the hot, low-density interstellar medium and the warm, partially ionized clouds, respectively. In each figure, spectra are shown for black holes at 3~pc and 15~pc, varying either $\dot{m}$ or $i$. Black holes in the hot interstellar medium remain undetectable even with a jet. For black holes in the warm clouds, however, the jet can significantly enhance detectability, particularly in the radio and X-ray bands. While the eROSITA detection threshold was far above the predicted X-ray emission in the pure ADAF case, the jet component makes the X-ray band a promising avenue for discovery.

We again emphasize that the jet model of \citet{2014MNRAS.442L.110X} is phenomenological. The spectra presented in Figs.~\ref{f-hot}–\ref{f-warm2} serve as illustrative examples, but their resemblance to real accretion spectra remains unknown, as no black hole accreting from the interstellar medium has yet been identified.


\section{Searching for Nearby Black Holes in Gaia~DR3}\label{s-gaia}

\subsection{Accreting Black Holes}

Although significant uncertainties remain regarding the local black hole population, the volume fraction occupied by Local Interstellar Clouds, and the precise electromagnetic spectrum of an isolated black hole accreting from the interstellar medium, the discussion in the preceding sections motivates a search for candidate accreting black holes in the Gaia~DR3 catalog. We also note that our analysis does not strictly require that such black holes be isolated. A nearby black hole in a binary system with another black hole or a neutron star could also be discovered by searching for objects accreting from the interstellar medium. Conversely, if a nearby black hole were in a binary system with a normal star, it would presumably have already been identified via the motion of the companion star. However, the probability of finding such a system within 15~pc of Earth is very low, as only about 0.16\% of stellar-mass black holes in the Galactic disk are paired with a normal star~\citep{2020A&A...638A..94O}.

We begin by selecting all Gaia DR3 sources within 15~pc of the Solar System, yielding 1,071 objects\footnote{We note that the assumption that distance is the inverse of parallax is only valid for high signal-to-noise ratio values of the parallax but can give highly biased results in general \citep{2021AJ....161..147B}. In our case, the high signal-to-noise ratio condition is met.}. We cross-match these sources with the SIMBAD database~\citep{2000A&AS..143....9W}\footnote{\href{https://simbad.u-strasbg.fr/simbad/}{https://simbad.u-strasbg.fr/simbad/}} to identify known objects and assess their classifications. Of these, five sources have no entry in SIMBAD, and one source is listed as a brown dwarf candidate; the remaining sources are well-classified in SIMBAD. Table~\ref{t-gaia} presents the source IDs and basic properties for the five SIMBAD-absent sources and the brown dwarf candidate. We note that these six sources do not appear in the unWISE catalog \citep{2019ApJS..240...30S}, in 2MASS~\citep{2003tmc..book.....C}, in the ``living'' catalog of nearby faint objects \citet{best_2025_15802304}, and in the list of new nearby sources in \citet{2026arXiv260109671V}.

\begin{figure*}[htpb]
\centering
\includegraphics[width=0.60\textwidth]{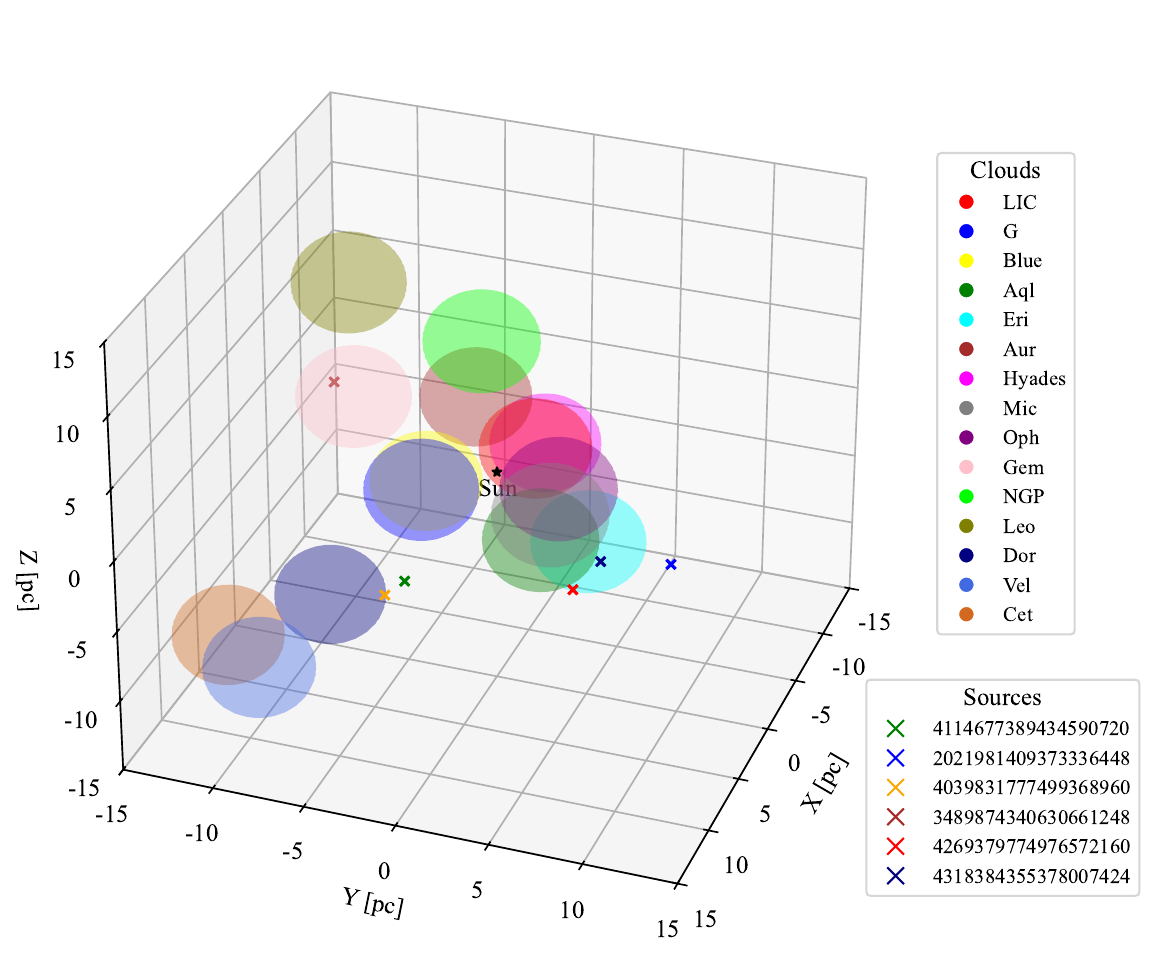} \\
\includegraphics[width=0.32\textwidth]{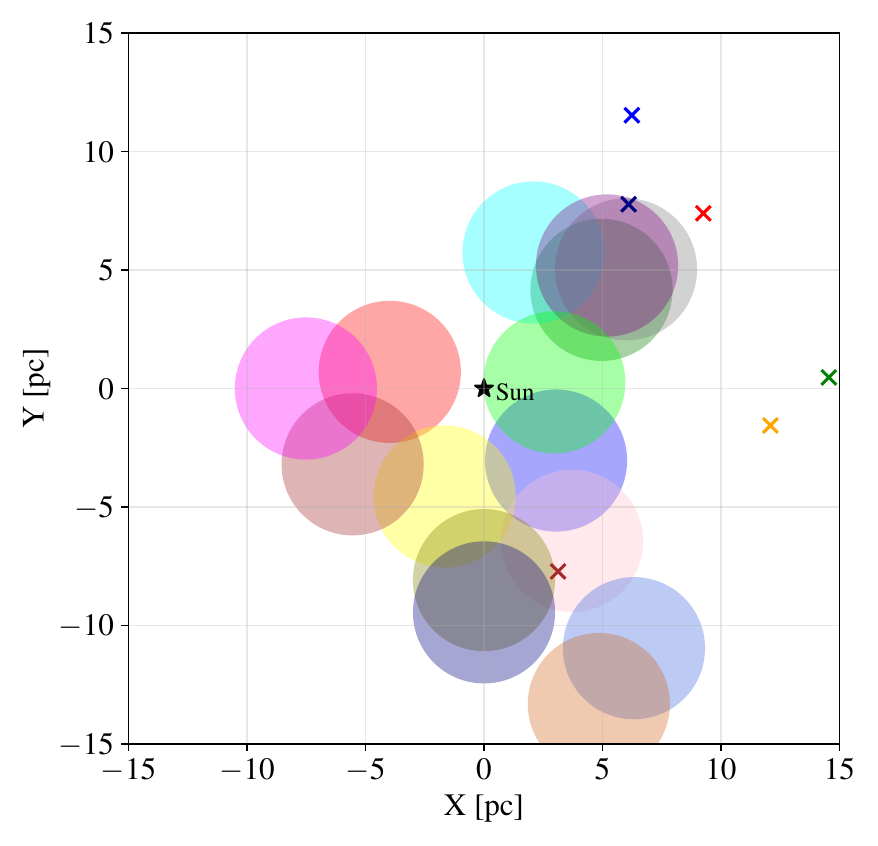}
\includegraphics[width=0.32\textwidth]{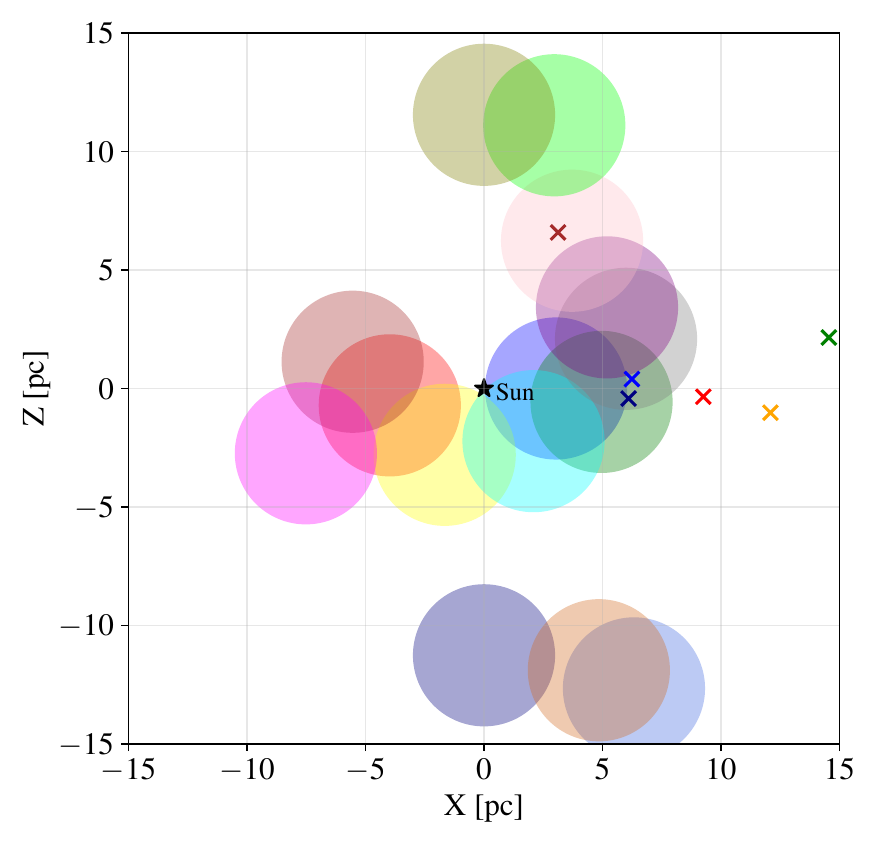}
\includegraphics[width=0.32\textwidth]{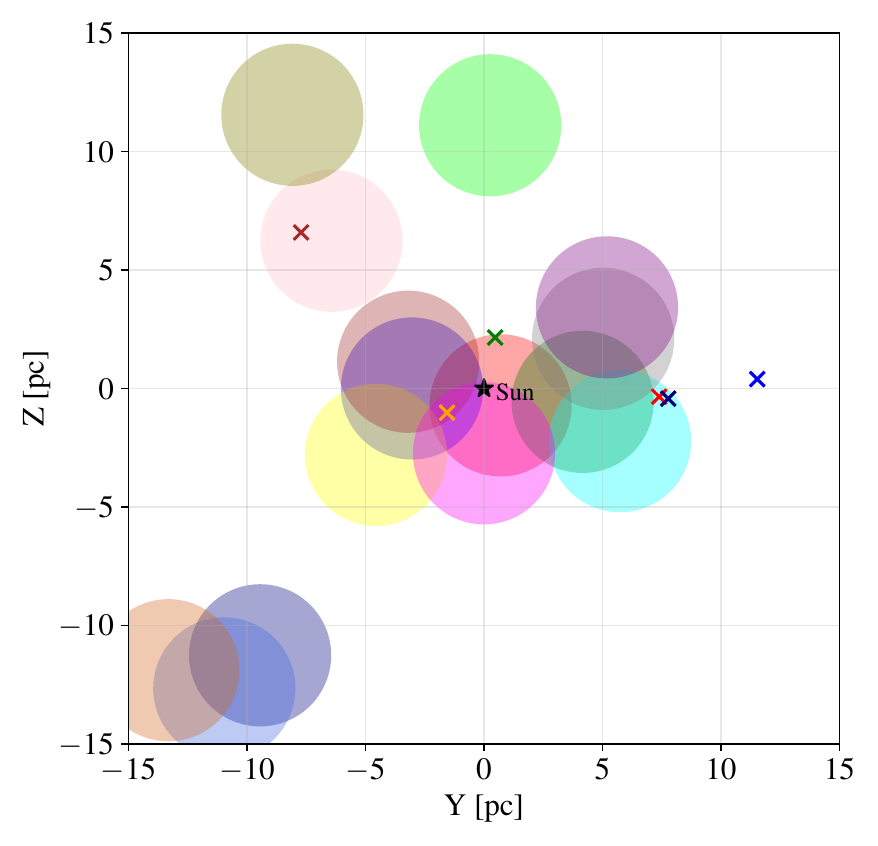}
\caption{Three-dimensional map (top) and projections onto the XY, XZ, and YZ planes (bottom) of the region within 15~pc of the Solar System, showing the locations of the 15~Local Interstellar Clouds and the six Gaia sources listed in Table~\ref{t-gaia}. Each cloud is represented by a sphere of radius 3~pc with a distinct color. The Solar System is at the center of the map, indicated by the black star. See the text for further details.}
\label{f-lism}
\end{figure*}

These six faint sources (all near the Gaia detection limit) are potential candidates for isolated black holes accreting from the interstellar medium. However, the brown dwarf candidate can be excluded as a stellar-mass black hole, as it resides in a binary system~\citep{2021RNAAS...5...40S}. The five sources not in SIMBAD all lie near the Galactic plane. While no further information is available from the public Gaia~DR3 catalog, their locations make it likely that they are spurious astrometric solutions: the Galactic plane is a region of crowding which is challenging for Gaia. Such sources may be removed in Gaia~DR4, scheduled for release in December 2026.

The three-dimensional map of the Solar neighborhood is shown in Fig.~\ref{f-lism}. The location of the Sun is indicated by the black star at the center of this region, while the locations of the six faint sources in Table~\ref{t-gaia} are marked by crosses of different colors. The warm, partially ionized clouds are modeled as spheres of radius 3~pc, which together fill 12\% of the local volume within 15~pc. Their positions are inferred from the measurements reported in \citet{2025ApJ...986...58Z}. We note that modeling all clouds as spheres of radius 3~pc is a very crude approximation and cannot provide definitive conclusions about the actual interstellar medium density around the six candidates\footnote{While the projections of these clouds on the sky are relatively well constrained observationally \citep[see][]{2008ApJ...673..283R}, there are large uncertainties in their exact positions and their thicknesses along our line of sight. The positions and the thicknesses of these clouds are determined by studying interstellar absorption in the spectra of nearby stars, so we can only infer lower limits on the distances of the cloud boundaries \citep{2025ApJ...986...58Z}.}. Under this simple model, only source 3489874340630661248 -- the brown dwarf candidate in SIMBAD -- lies inside one of the clouds, while the other five sources are outside. If we instead assume the clouds are spheres of radius 3.5~pc, source 4318384355378007424 also falls inside a cloud.

\begin{figure}[htpb]
\centering
\includegraphics[width=0.45\textwidth]{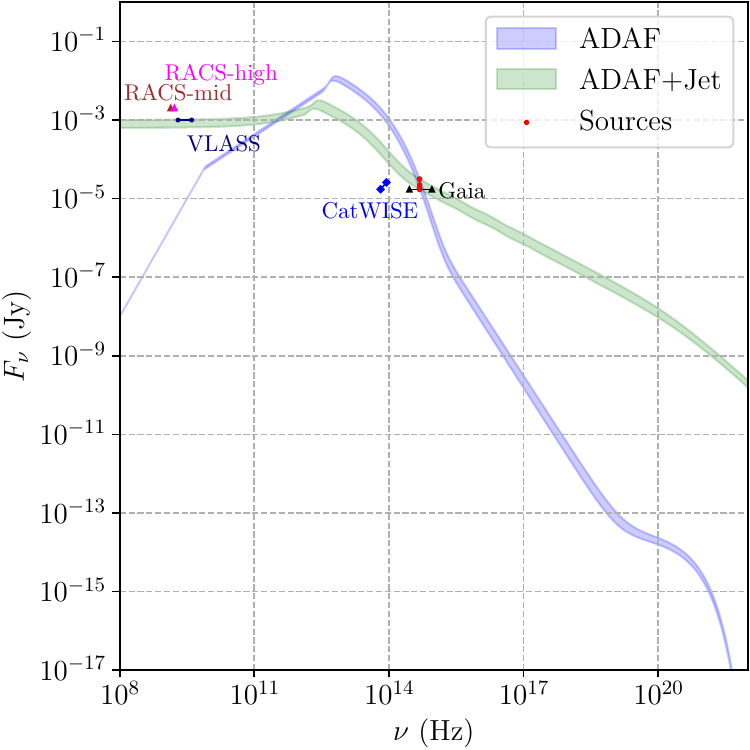}
\caption{Predicted spectra of the six Gaia sources in Table~\ref{t-gaia} under the ADAF model (Subsection~\ref{ss-adaf}) and the ADAF+Jet model (Subsection~\ref{ss-adaf+jet}). Also shown are the detection thresholds of the Gaia~DR3, CatWISE, VLASS~QL1, RACS-mid, and RACS-high catalogs.}
\label{f-spectra}
\end{figure}

\begin{figure*}[htpb]
\centering
\includegraphics[width=0.37\textwidth]{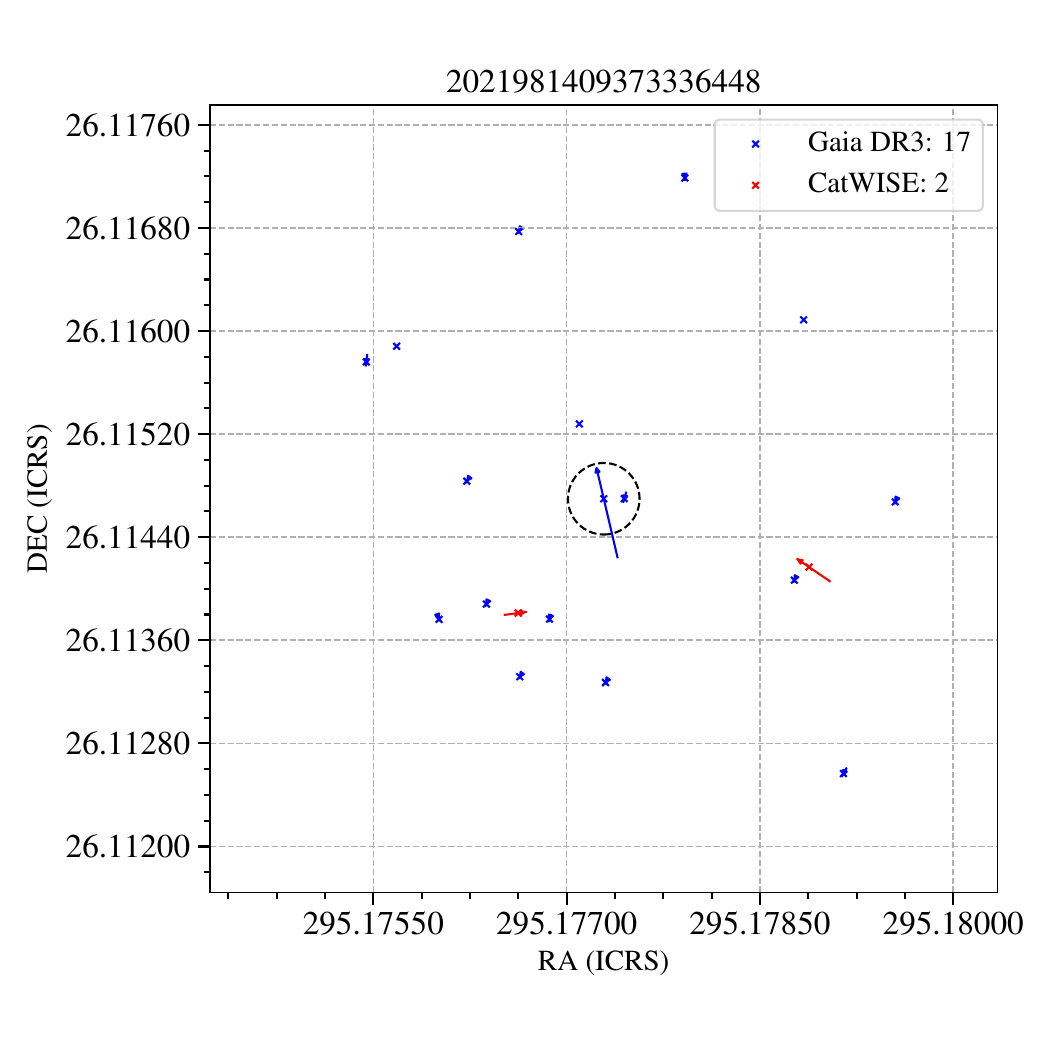}
\hspace{0.8cm}
\includegraphics[width=0.37\textwidth]{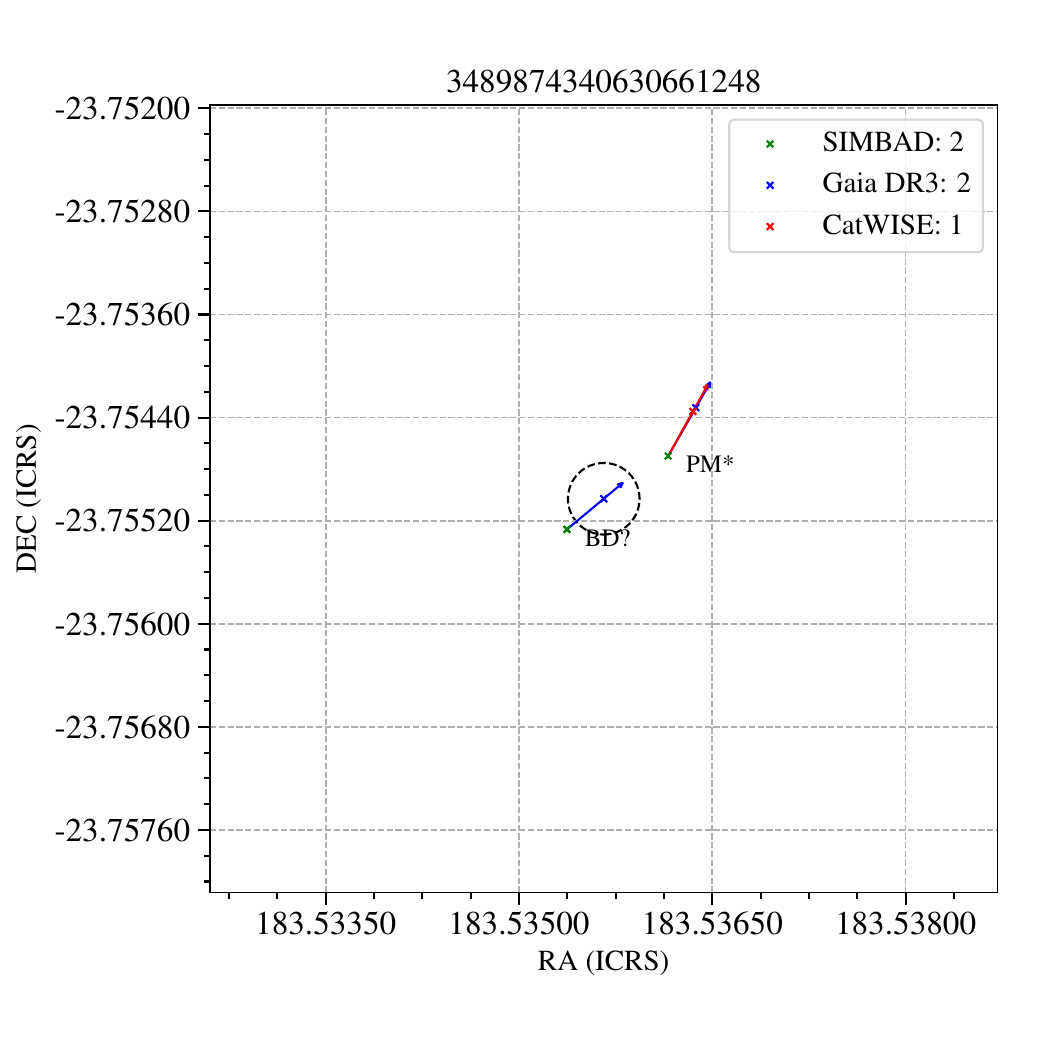} \\
\includegraphics[width=0.37\textwidth]{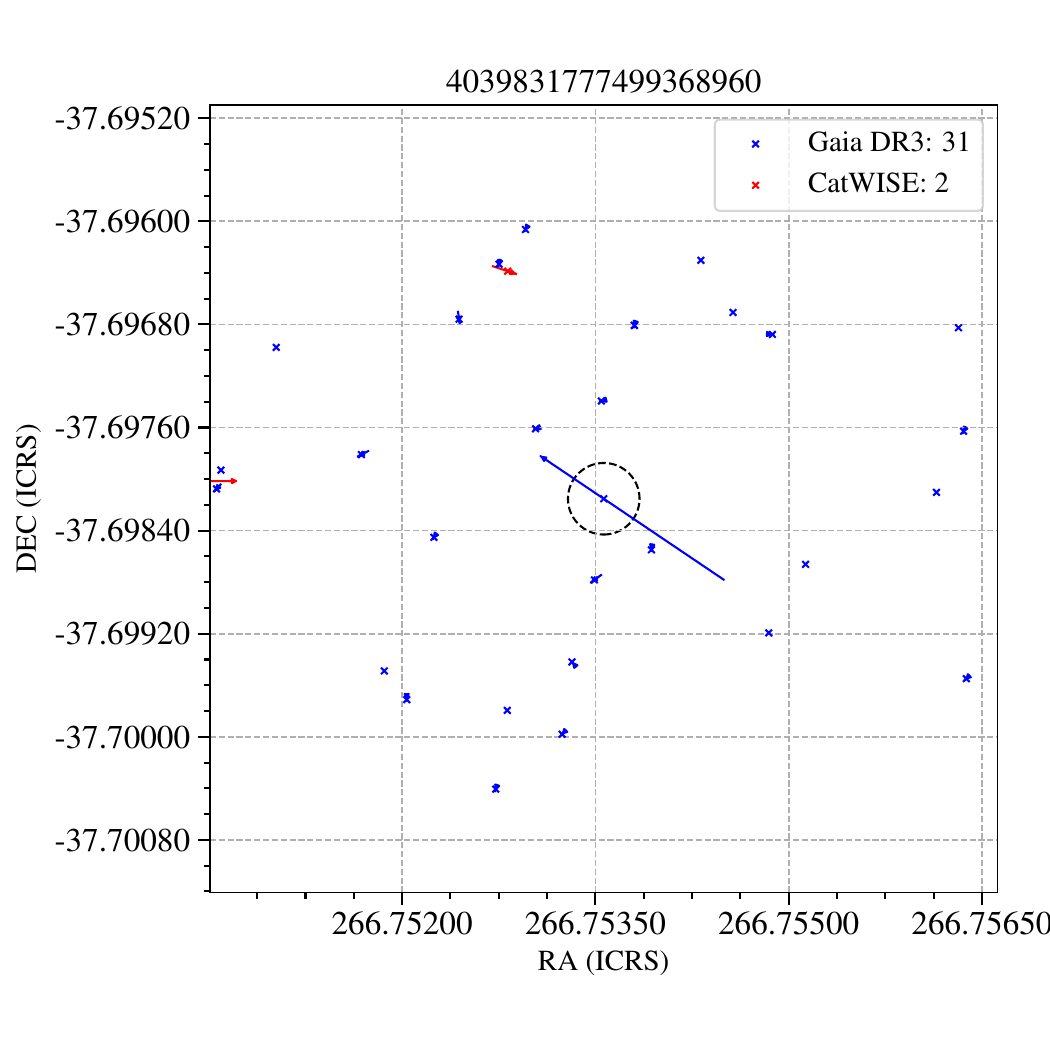}
\hspace{0.8cm}
\includegraphics[width=0.37\textwidth]{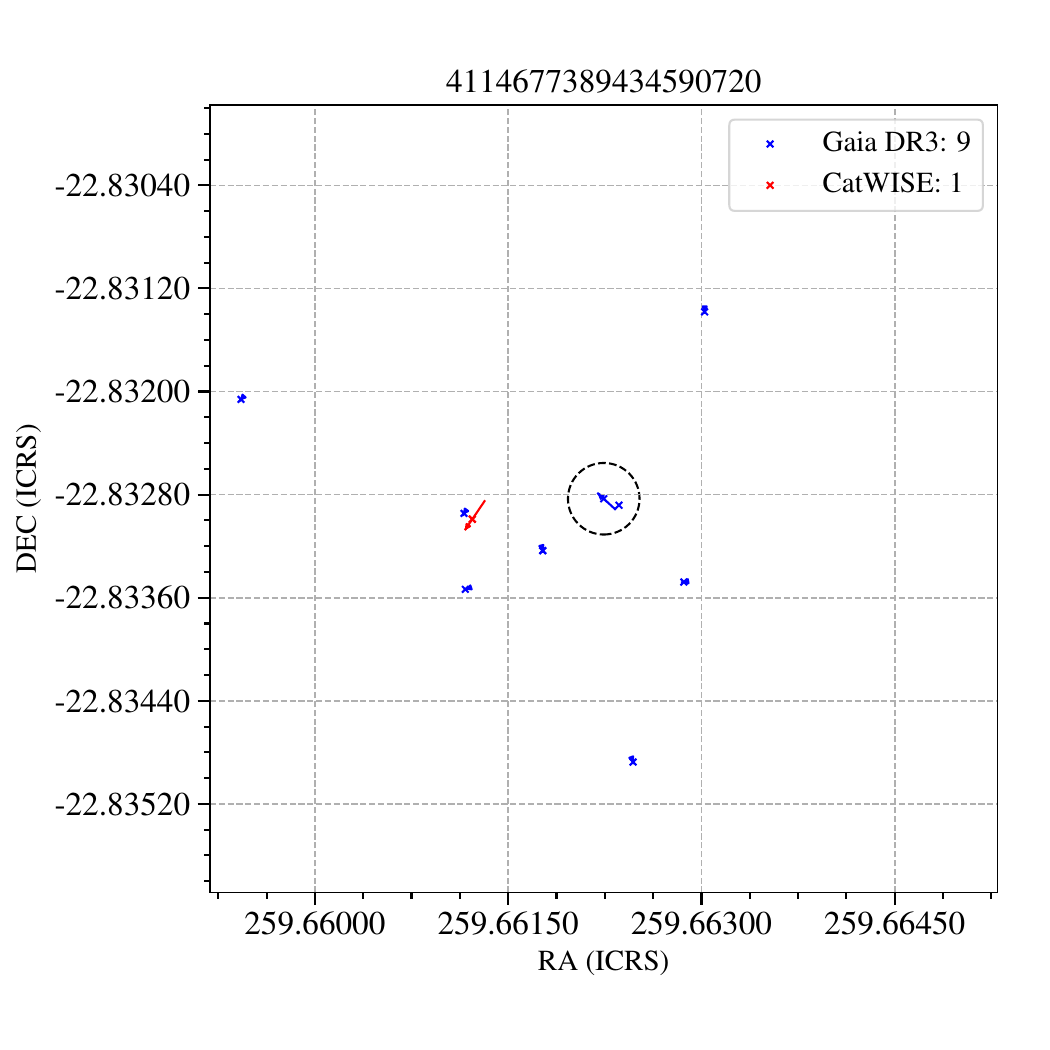} \\
\includegraphics[width=0.37\textwidth]{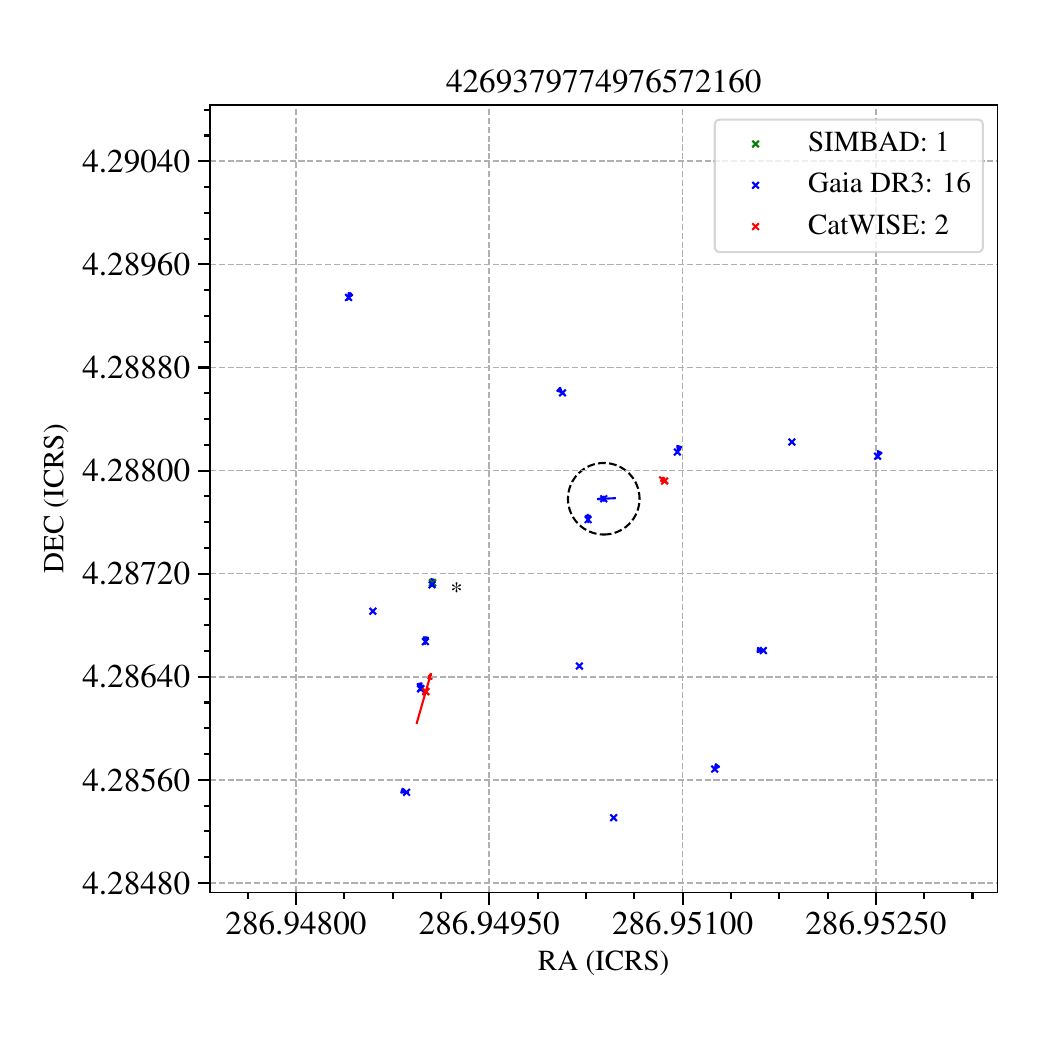}
\hspace{0.8cm}
\includegraphics[width=0.37\textwidth]{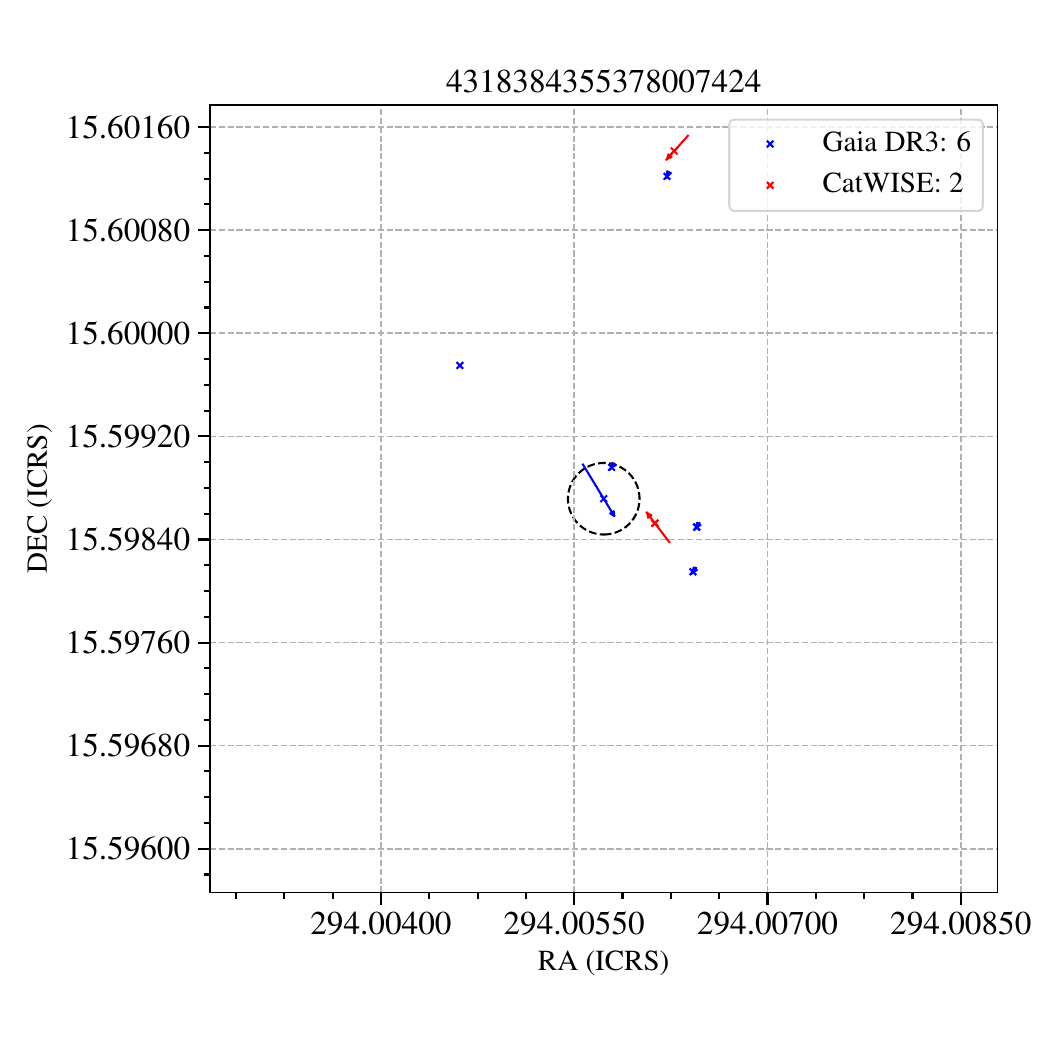}
\vspace{-0.3cm}
\caption{Maps of the regions around the six Gaia sources in Table~\ref{t-gaia}. Each map shows a 10~as radius region centered on one of the six sources, with all sources from the SIMBAD, Gaia~DR3, CatWISE, VLASS~QL1, RACS-mid, and RACS-high catalogs overplotted. Catalogs that have no sources in a given map are omitted from the legend. VLASS~QL1, RACS-mid, and RACS-high never appear in the legend, as no radio sources from these catalogs are found around any of the six Gaia sources. For sources with proper motion measurements in CatWISE or Gaia~DR3, the trajectories from 2000 to 2025 are shown, with the arrow indicating the source position in 2025 and the source plotted at its position at the epoch of observation. The black dashed circle at the center of each map indicates a radius of 1~as.}
\label{f-sources}
\end{figure*}

If our six candidates are stellar-mass black holes accreting from the interstellar medium, we can use the models presented in Subsections~\ref{ss-adaf} and~\ref{ss-adaf+jet} to predict their spectra from the radio band to the $\gamma$-ray band. The results are shown in Fig.~\ref{f-spectra}, where we also report the detection thresholds of the infrared catalog CatWISE~\citep{2021ApJS..253....8M} and the radio catalogs VLASS~\citep{2021ApJS..255...30G} and RACS~\citep{2024PASA...41....3D,2025PASA...42...38D}\footnote{CatWISE covers the full sky. VLASS covers approximately 80\% of the sky ($\delta > -40^\circ$) and here we use VLASS QL (Quick Look) 1. RACS comprises three catalogs: RACS-low (887.5~MHz), RACS-mid (1367.5~MHz), and RACS-high (1655.5~MHz). We omit RACS-low due to its poor angular resolution. The RACS catalogs cover about 80\% of the sky ($\delta < 49^\circ$) and are complementary to VLASS.}. We searched these catalogs for radio and infrared emission associated with our candidates, with negative results. Fig.~\ref{f-sources} shows the sky regions around our six sources, along with all sources in those regions from SIMBAD, CatWISE, VLASS~QL1, RACS-mid, and RACS-high (no sources from VLASS~QL1, RACS-mid, or RACS-high are present in these regions). Although the detection thresholds of the radio catalogs are close to the fluxes predicted by our models (see Fig.~\ref{f-spectra}), both models agree that, if these sources were black holes accreting from the interstellar medium, they should be detectable in CatWISE. The observation periods of the two catalogs overlap (2010-2018 for CatWISE and 2013-2025 for Gaia), ruling out the possibility that these sources were too faint during the CatWISE observations but bright enough during the Gaia observations.

We note that Gaia~DR3 contains sources without parallax measurements\footnote{Gaia DR3 includes sources with two (position only), five (position, parallax, proper motion), or six (adding source color) astrometric parameters. Sources with insufficient observations or poor astrometric fits lack a measured parallax.}. Many are faint and could be of interest, but their distances cannot be determined from the public catalog. Parallaxes for some may become available in Gaia~DR4.

\subsection{Non-Accreting Black Holes}

Indirect detection of an isolated black hole within 15~pc through its gravitational influence on nearby stars is highly unlikely due to the low local stellar density.

The stellar velocity dispersion in the Solar neighborhood is $\sigma \sim 20$~km~s$^{-1}$. A close encounter between a star and a black hole could produce a high-velocity star. The radius of influence of the black hole is approximately $R \sim 2 G_{\rm N} M_{\rm BH}/\sigma^2 \sim 0.2$~mpc. With $\sim 10^3$ stars within 15~pc (stellar density $n_{\rm stars} \sim 0.08$~pc$^{-3}$), the probability of such an encounter per black hole per year is
\begin{eqnarray}
\Gamma \sim n_{\rm stars} \, \sigma \, \pi R^2 \sim 2 \times 10^{-13}~{\rm yr}^{-1}~{\rm BH}^{-1}.
\nonumber
\end{eqnarray}
Assuming 10 black holes within this volume, the total encounter rate is $\sim 10^{-12}$~yr$^{-1}$, which is far too low for a detectable event within any plausible observational timescale.

\section{Concluding Remarks}\label{s-conclusions}

It is plausible that one or more stellar-mass black holes reside within 15~pc of the Solar System. Most stellar-mass black holes are expected to be isolated, and the majority of those in binary systems likely have another black hole as a companion. This makes their direct detection particularly challenging.

If such a black hole is located within one of the Local Interstellar Clouds -- which collectively occupy 5-20\% of the local volume -- accretion from the interstellar medium could produce a detectable electromagnetic signal. However, substantial uncertainties in the expected accretion spectra complicate the formulation of a definitive observational strategy. For a simple advection-dominated accretion flow (ADAF) spectrum, detection may be more feasible in the radio to optical bands. If the spectrum is jet-dominated, wide-area X-ray surveys could prove more effective for identifying these faint sources. Conversely, black holes outside these clouds, surrounded by the hot, low-density interstellar medium, have accretion luminosities too low for detection with current or near-future facilities, regardless of the accretion model.

Motivated by the possibility of a black hole within one of the Local Interstellar Clouds, we searched the Gaia~DR3 catalog for candidate objects. Of the 1,071 sources within 15~pc with measured parallaxes, only six lack a clear classification: five are absent from SIMBAD, and one is listed as a brown dwarf candidate. The latter can be excluded as a stellar-mass black hole because it resides in a binary system. The five SIMBAD-absent sources all lie near the Galactic plane, making them likely spurious astrometric solutions, for instance caused by unmodelled background sources (crowding) and/or unmodelled binarity, though their nature cannot be definitively ascertained from the public catalog alone. Our failure to detect infrared emission from these sources in CatWISE, or radio emission in VLASS and RACS, reinforces the conclusion that they are unlikely to be accreting black holes.

Constructing a physically motivated list of expected Gaia properties to aid candidate selection is unfortunately not feasible. As noted by \citet{2025ApJ...988L..12M}, an isolated black hole accreting from the interstellar medium would appear as a faint source with a relatively featureless spectrum. Consequently, identification cannot rely on single-instrument observations. Multi-wavelength follow-up campaigns, covering radio, optical, and X-ray bands, are necessary to rule out alternative interpretations and confirm the nature of any candidate.

Future efforts will focus on searching for isolated black holes accreting from the warm, partially ionized interstellar medium in radio and X-ray catalogs. The forthcoming release of Square Kilometre Array (SKA) data, currently anticipated in the first half of 2027, appears especially promising for this endeavor.


\vspace{0.5cm}

{\bf Acknowledgments --}
This work was supported by the National Natural Science Foundation of China (NSFC), Grant No.~W2531002, and the Fudan-Warwick Joint Seed Fund, Grant No.~JMH6282518.
A.N. also acknowledges the support from the Shanghai Government Scholarship (SGS).
F.G.X. is supported in part by the National Natural Science Foundation of China (NSFC), Grant No.~12373017, and the State Key Laboratory of Radio Astronomy and Technology (Chinese Academy of Sciences).
This work has made use of data from the European Space Agency (ESA) mission {\it Gaia} (\url{https://www.cosmos.esa.int/gaia}), processed by the {\it Gaia} Data Processing and Analysis Consortium (DPAC, \url{https://www.cosmos.esa.int/web/gaia/dpac/consortium}). Funding for the DPAC has been provided by national institutions, in particular the institutions participating in the {\it Gaia} Multilateral Agreement. 
This research has made use of the SIMBAD database, operated at CDS, Strasbourg, France.

\bibliographystyle{aa}
\bibliography{references}

\end{document}